# Stereo Speech Enhancement Using Custom Mid-Side Signals and Monaural Processing

**AARON S. MASTER, LIE LU, AND NATHAN SWEDLOW**
(amaster@gmail.com, Lie.Lu@dolby.com, Nathan.Swedlow@dolby.com)

*Dolby Laboratories, Inc, 1275 Market St, San Francisco, CA 94103, United States*

Speech Enhancement (SE) systems typically operate on monaural input and are used for applications including voice communications and capture cleanup for user generated content. Recent advancements and changes in the devices used for these applications are likely to lead to an increase in the amount of two-channel content for the same applications. However, SE systems are typically designed for monaural input; stereo results produced using trivial methods such as channel independent or mid-side processing may be unsatisfactory, including substantial speech distortions. To address this, we propose a system which creates a novel representation of stereo signals called Custom Mid-Side Signals (CMSS). CMSS allow benefits of mid-side signals for center-panned speech to be extended to a much larger class of input signals. This in turn allows any existing monaural SE system to operate as an efficient stereo system by processing the custom mid signal. We describe how the parameters needed for CMSS can be efficiently estimated by a component of the spatio-level filtering source separation system. Subjective listening using state-of-the-art deep learning-based SE systems on stereo content with various speech mixing styles shows that CMSS processing leads to improved speech quality at approximately half the cost of channel-independent processing.

## 0 INTRODUCTION

Speech enhancement (SE) is a technology which aims to reduce or eliminate background noise while preserving speech quality, typically for monaural audio signals in applications including voice communications and capture clean-up. State-of-the-art deep learning-based SE systems include FullSubnet [1], DPCRN [2], and NSNet2 [3] which is the official baseline for the DNS challenge [4]. These technologies are relatively mature for monaural input but are not designed for stereo (or higher channel count) inputs; pilot testing of these systems using channel-independent processing of stereo signals with various styles of speech mixing found that they tend to produce unsatisfactory outputs with significant noise leakage and speech distortion. For channel independent processing, the SE systems are likely to exhibit unequal strain versus time and frequency in each channel, and the unequal imperfections may be perceived as additional distortions.

Due to recent trends in hardware and operating systems, there is likely to be a large increase in the amount of stereo data available for SE applications. One major consumer electronics company recently updated their operating system to allow access to both mics on their mobile phones and tablets [5]; the company has over 1.8 billion active devices [6]. An operating system provider has facilitated the inclusion of microphone arrays (including two-mic arrays) on laptops for several years [7]; it is possible they will also facilitate accessing array signals directly. Other advancements include binaural microphones on headphones (e.g., [8, 9]) and an increase in the availability of affordable peripheral stereo microphones.

This likely increase in the availability of stereo inputs to SE systems which are not designed for such inputs presents an opportunity for improvement. In this paper, we present a system which allows an existing monaural SE system to process a new type of signal based on a spatially dynamic version of mid-side signals. To develop an understanding of the new signal, we will describe the special relationship between center-panned speech target sources and standard mid-side signals, namely that the mid signal boosts the target source relative to other signals, while the side signal suppresses it. We will then describe how we may generalize this concept to target sources which are *not* center-panned, whose mixing may be estimated using a technique described in [10]. This allows creation of *custom mid-side signals* (CMSS) which allow arbitrarily mixed speech to receive the same benefit as center-panned speech in standard mid-side signals. A SE system may process the custom mid signal, instead of processing standard mid, side or channel signals. Using a subjective listening test, we demonstrate that for state-of-the-art deep-learning based SE systems, this approach allows for significantly improved speech output quality at approximately half the processing cost of channel-independent processing.

This paper is organized as follows. In section 1, we review mid-side signals and describe how they can be generalized and customized using detectable mixing parameters to create *custom mid-side signals* (CMSS). In section 2, we describe how to use CMSS for efficient, high-quality processing of stereo input by a monaural SE system. In section 3, we present results from a subjective listening test of speech enhancement systems, which compares the perceived qualities of the proposed method with those of a channel-independent baseline. We conclude in section 4 with a summary and discussion of future work.

## 1 MID-SIDE SIGNALS

In this section, we describe how CMSS are created and motivate their use. To do so, we begin in the first subsection by reviewing standard mid-side signals and note their special relationship with center-panned sources. In the second subsection, we describe a more general target source mixing model, which uses additional parameters $\Theta_1$ and $\Phi_1$ to describe mixing of sources which are *not* necessarily center-panned. In the third subsection, we describe how to create generalized mid-side signals which provide mid-side benefits to sources which are mixed using the more general model. In the fourth subsection, we describe how the concept of generalized mid-side signals may be effectively implemented for real world signals by dynamically estimating $\Theta_1$ and $\Phi_1$ over time and frequency using a technique in [10]. We term the resulting signals CMSS.

### 1.1 Standard Mid-Side

We presently summarize the standard mid-side signal decomposition (see, e.g., [11]). We do not presently describe mid-side microphone capture, though we note that a mid-side signal decomposition of stereo signals may approximately recover the components signals in a mid-side recording; see, e.g., [12]. The standard mid-side decomposition of a stereo signal is as follows:

$$\begin{aligned} M &= 0.5(L + R) \\ S &= 0.5(L - R) \end{aligned} \quad (1)$$

where $M$ is the mid signal, $S$ is the side signal, $L$ is the left channel signal and $R$ is the right channel signal. As the operations used here are linear, this calculation may occur in the time domain or a time-frequency domain such as the short time Fourier transform (STFT) domain; below we will work in the STFT domain. We may recover the original stereo channels via:

$$\begin{aligned} L &= M + S \\ R &= M - S. \end{aligned} \quad (2)$$

For inputs which contain a center-panned target signal of interest, the mid signal will contain the target signal (and likely, other sounds) while the side signal will be devoid of the target signal but will contain other non-center-panned sounds, depending on how they are mixed to the two channels. (See, e.g., sec. 4.4 of [13].) For stereo signals with a center-panned target signal, the mid-side representation may be thought of as providing mild source separation or source boosting; the mid signal will increase the relative level of center-panned in-phase signal components (and of components which are approximately so) while attenuating others; the side signal will completely attenuate center-panned signal components and will boost out-of-phase components.

This special relationship between a center-panned target source and the mid-side signals can be beneficial to a processing system which seeks to enhance a target source, as we will describe below. In the subsections that follow, we will describe how standard mid-side signals can be generalized and customized to extend this benefit to a much larger class of stereo input signals. In order to do so, we next introduce a more general source mixing model which describes target sources which are not necessarily center-panned.

### 1.2 Generalized Mixing Model

In order to develop a more generalized version of mid-side signals, we must develop a more generalized mixing model for the target source. This subsection describes such a model; in the next subsection we will then develop a generalized version of mid-side signals based on this model.

Our model considers how a monaural source $S_1$ with magnitude $|S_1|$ and phase $\Psi_1$ for each STFT tile $(\omega, t)$ is mixed to two channels ($L$ and $R$) in STFT space. First, we define the mono source as

$$S_1(\omega, t) = |S_1(\omega, t)| \exp(i\, \Psi_1(\omega, t)) \quad (3)$$

where we note that values of $S_1$ exist for each STFT tile of frequency bin $\omega$ and frame $t$; going forward we shall abbreviate $S_1(\omega, t)$ (and similar such quantities) as $S_1$ (and similar) for simplicity.

We shall model the mixing with regard to interchannel level difference (ILD) and interchannel phase difference (IPD). For purposes of developing generalized mid-side equations, we temporarily treat each of these quantities as a single fixed value for all times and frequencies. (In practice, these quantities will vary, and it is critical to allow the IPD to vary vs frequency and time in order to model sources mixed with reverberation or interchannel delay; see Sec. 3 of [10]). The ILD is modeled by a panning coefficient $\Theta_1$ ranging from 0 (pure left) to $\pi/2$ (pure right) under the constant power panning law [14], leading to the following values of $S_1$ in the $L$ and $R$ channels when the IPD is zero:

$$\begin{aligned}L &= S_1 \cos(\Theta_1) \\ R &= S_1 \sin(\Theta_1).\end{aligned} \quad (4)$$

The IPD is described by the parameter $\Phi_1$. When we model mixing with nonzero IPD, we must explicitly define the relationship between $\Psi_1$ and $\Phi_1$. One option [15] is to declare the left channel phase to be the true source phase, in which case the phase difference applies only to the right channel. However, doing this creates a problem when describing source phase and IPD for an extreme right panned source. In such cases, the left channel's phase information is unrelated to the target source and effectively random, influenced by values in the noise floor or backgrounds. Modeling the IPD as split evenly between the channels creates a similar problem; the left channel is still random. To address this, we use a mixing model where the channel in which the source is stronger in power has proportionally closer phase to the source phase $\Psi_1$ and the other channel is proportionally less close to $\Psi_1$, and dictated by $\Phi_1$. By careful selection of these proportions based on $\Theta_1$ data, we can also ensure that the phase difference between the channels still equals $\Phi_1$. The $L$ and $R$ channels are thus modeled as:

$$\begin{aligned}L &= S_1 \cos(\Theta_1) \exp(i\, \Phi_1 \sin^2 \Theta_1) \\ &= |S_1| \exp(i\, \Psi_1) \cos \Theta_1 \exp(i\, \Phi_1 \sin^2 \Theta_1) \\ &= |S_1| \cos \Theta_1 \exp(i\, (\Psi_1 + \Phi_1 \sin^2 \Theta_1)) \\ R &= S_1 \sin(\Theta_1) \exp(-i\, \Phi_1 \cos^2 \Theta_1) \\ &= |S_1| \exp(i\, \Psi_1) \sin(\Theta_1) \exp(-i\, \Phi_1 \cos^2 \Theta_1) \\ &= |S_1| \sin(\Theta_1) \exp(i\, (\Psi_1 - \Phi_1 \cos^2 \Theta_1)).\end{aligned} \quad (5)$$

We see from examining equations (5) that, for tiles in which only the target source is present, the IPD, calculated via $\angle(L/R)$, (see [10]) will equal $\Phi_1$. Similarly, the ILD, calculated via $\arctan(R/L)$ will equal $\Theta_1$. An appendix further explores these calculations and their relationship with $|S_1|$ and $\Psi_1$.

### 1.3 Generalized Mid-Side

Now that we have described a generalized target source mixing model, we may describe generalized mid-side signals based on this model. It can be seen that for a target source mixed as specified in the equations (5) above, we can define generalized mid and side signals which will have the boosting and elimination properties that standard mid-side signals have for center-panned signals. Equations (6) below specify such signals, which we term *generalized mid-side signals*. (For a more detailed derivation of the side signal, see "normalized weighted subtraction" on p. 38 of [13]; the derivation of the mid signal is similar.)

$$\begin{aligned} M &= c_1 L + c_2 R \\ S &= c_3 L + c_4 R \end{aligned} \quad (6)$$

where

$$\begin{aligned} c_1 &= \cos\Theta_1 \, \exp(-i\Phi_1 \sin^2\Theta_1) \\ c_2 &= \sin\Theta_1 \, \exp(i\Phi_1 \cos^2\Theta_1) \\ c_3 &= \sin\Theta_1 \, \exp(-i\Phi_1 \sin^2\Theta_1) \\ c_4 &= \cos\Theta_1 \, \exp(i\Phi_1 \cos^2\Theta_1). \end{aligned}$$

The *inversion* equations (or *stereo reconstruction equations*), which return to conventional stereo signals are:

$$\begin{aligned} L &= (M\cos\Theta_1 + S\sin\Theta_1)\exp(i\Phi_1\sin^2\Theta_1) \\ R &= (M\sin\Theta_1 - S\cos\Theta_1)\exp(-i\Phi_1\cos^2\Theta_1) \end{aligned}. \quad (7)$$

As a point of clarification, we note that for a center-panned target source (for which $\Theta_1 = \pi/4$ and $\Phi_1 = 0$) these generalized mid-side equations (and thus the inversion equations) use different scaling than the standard mid-side equations but are otherwise identical. The inversion equations still recover $L$ and $R$ channels with the original scale intact. Using the noted values of $\Theta_1$ and $\Phi_1$ leads to:

$$\begin{aligned} M &= \sqrt{2}/2 \, (L+R) \\ S &= \sqrt{2}/2 \, (L-R) \\ L &= \sqrt{2}/2 \, (M+S) \\ R &= \sqrt{2}/2 \, (M-S). \end{aligned} \quad (8)$$

**1.4 Custom Mid-Side**

We can further expand the generalized mid-side signal concept of the previous subsection by allowing the parameters $\Theta_1$ and $\Phi_1$ to vary versus time and frequency. This effectively allows for special mid and side signals which track target source signals whose mixing parameters are dynamic. An example is a target source which moves (relative to stereo capture microphones) whose corresponding $\Theta_1$ and $\Phi_1$ values will vary with time. (And as noted above, $\Phi_1$ must be allowed to vary with time and frequency to model reverberant sources or those mixed with interchannel delay.) We term the time- and frequency-varying mid-side signals *custom mid-side signals* (CMSS). Their mathematical expressions are identical to those introduced in the previous subsection, with additional flexibility in that $\Theta_1$ and $\Phi_1$ are allowed to vary by frequency sub-band $b$ and time frame $t$; they become $\Theta_1(b,t)$ and $\Phi_1(b,t)$.

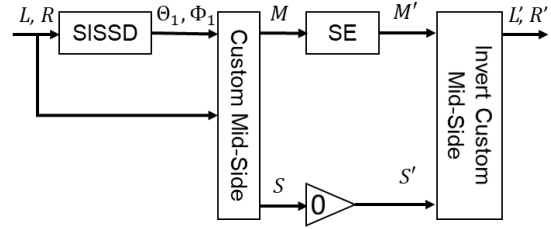

Fig. 1: Stereo Speech Enhancement using CMSS.

In order to generate CMSS we must estimate and apply $\Theta_1(b,t)$ and $\Phi_1(b,t)$ which we term *dynamic mixing parameters*. To estimate these parameters we use a process described in [10] termed *spatially identifiable subband source detection* or SISSD (and where the parameters are also termed *thetaMiddle* and *phiMiddle*). The parameters are estimated at a frequency granularity of one parameter per quasi-octave frequency sub-band (sub-band edges are [0, 400, 800, 1600, 3200, 6400, 13200, 24000] Hz) and updated once every 1024 samples for 48 kHz sampled audio. The choice of granularity is critical; if we choose very large frequency bands, reverberant sources cannot be well estimated, and if we choose to update values infrequently vs time, then rapidly moving sources cannot be tracked. However, choosing too fine a granularity leads to unreliable and unstable estimates. In the most extreme case, one could estimate $\Theta_1$ and $\Phi_1$ for each STFT tile which leads to the mid signal containing all energy and the side signal containing none. In that case, the parameters do not characterize meaningful target source mixing but rather the individual statistics of a single tile or micro-region. The choice of granularity is further explored in sec. 2 of [10].

We also note that the model used here characterizes the mixing of a monaural source to two channels using only the two (time- and frequency-varying) parameters $\Theta_1(b,t)$ and $\Phi_1(b,t)$. In practice, some sources, especially those mixed with heavy reverberation, cannot be so simply characterized. In such instances, the model used here effectively becomes a single wavefront approximation of the source mixing (see, e.g. appendix A of [13]), which we find still leads to a large quality improvement over existing baselines.

## 2 CMSS AND SPEECH ENHANCEMENT

### 2.1 Benefits of CMSS to SE

To summarize the above discussion, we have now greatly expanded the types of signals which can have a special relationship with a mid-side representation. A center-panned target source is boosted in a standard mid signal and eliminated from a standard side signal. With CMSS, however, any spatially concentrated target source detected by SISSD is boosted in the custom mid signal and eliminated or attenuated in the custom side signal.

We now revisit why this is beneficial. Knowing that a mid signal will contain the target source while the side signal will suppress it means that a SE system could process *only* the mid signal and still capture the target source. However, for standard mid-side signals, this property only holds if the target source is actually center-panned. If the speech is captured or mixed with interchannel delay or is at a higher level in one channel than the other, and the standard mid signal is used regardless, the mid signal may actually have a lower SNR than the side signal or channel signals. If an SE system were to process only the standard mid signal for such a case, it would lead to a result in which speech was underestimated or even entirely missed. By using SISSD to calculate parameters for CMSS, we can obtain a special, robust mid signal which boosts a target source while the special side signal eliminates or attenuates it. A SE system can then process only the custom mid signal before reconstructing the output.

### 2.2 Proposed Signal Flow and SE

Given the above description of how CMSS are obtained, we now describe a stereo processing method, depicted in Fig. 1, which can utilize any monaural SE system. We observe that the input signal enters as a stereo pair $(L, R)$ which the SISSD processes to obtain $\Theta_1(b,t)$ and $\Phi_1(b,t)$, which, along with the stereo input signal are passed to the CMSS generator. The generator produces the CMSS as described by equations (6) above. For convenience going forward, we shall term the *custom mid signal* only "CMS," noting that it does not include the side signal. The CMS is passed to a monaural speech enhancement system, whose output we term $M'$. The side signal is set to zero, and we call this signal $S'$. Finally, $M'$ and $S'$ are passed to a module which calculates $L'$ and $R'$ according to the CMSS reconstruction equations (7), thereby producing the system output. We shall term this output the CMS processed version; these signals are the ones evaluated as the CMS condition in the evaluation section.

In principle, any monaural SE system could be used in the overall system design proposed. For evaluation, we use two such systems, one developed internally, and the other an available state-of-the-art system. The internal system, which we term U-NetFB, is a SE network with a U-Net type architecture, similar in concept to [16, 17, 18, 19] but where the inputs are frequency band energies, rather than STFT bin values, and the outputs are real-valued frequency band softmask values. For the second system, we choose DPCRN [2] based on its relatively high performance compared with other state-of-the-art systems in a separate pilot test. Each of these systems is deep learning-based, and of much greater computational expense than the SISSD used to estimate $\Theta_1(b,t)$ and $\Phi_1(b,t)$. As a result, the computational cost of the proposed system is similar to a single instance of either SE system without the SISSD; adding a second instance

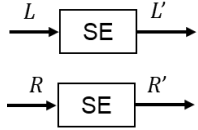

Fig. 2: Channel-Independent SE Processing

of an SE system, however, approximately doubles the cost.

### 2.3 Alternative Processing Options

We have proposed an overall system in which the SE component processes CMS only. However, there are alternatives which we now consider, along with their potential benefits and drawbacks. We will propose one such method, channel-independent processing, as a baseline for comparison with CMS processing. We will also describe why other options were not used as a proposed system or baseline.

*CMSS*. First we consider the system proposed in the previous subsection, but where the custom side signal is *also* processed by a second instance of SE rather than set to zero. This option was initially considered as it is more spatially exhaustive than the proposed method: processing both custom mid and side signals ensures that if speech is present, some version of it will be processed. However, testing of various real-world signals (including stereo-captured content and professionally generated stereo content) found that it is relatively rare for the custom side signal to contain significant undistorted speech energy compared with the custom mid signal. Processing a signal which contains little to no actual speech risks that speech will be erroneously detected, leading to perceptible errors. As noted above, an extra instance of SE also approximately doubles the computational cost. Given the cost and risks, this option was declined for now. Nonetheless, we consider this an area for future work as there are likely to be some signals, namely those for which the SISSD performs imperfectly, which benefit from CMSS processing, versus CMS-only processing.

*Standard mid-only*. Another variation on the proposed system is to process only the mid signal, but for a standard mid-side decomposition. For center-panned speech signals, this will have similar performance as CMS processing, while for other types of mixing, speech will be attenuated or missed entirely as noted above. For this reason, this option was not considered as a viable alternative or meaningful baseline.

*Standard mid-side*. A related idea is to process both standard mid and side signals for a given input, as they are similarly spatially exhaustive. For center-panned speech mixing, this approach has similar risks and costs (two SE instances) compared with CMSS processing, but for other speech mixing, this approach is similar in performance to channel-independent processing, as the standard mid-side signal pair may be understood as a spatial rotation of the original stereo signal pairs (see, e.g. Sec. 4.4 of [13]). Given the costs and risks, we declined this option.

*Alternative center*. Another novel option is to create an *alternative center* stereo signal from the CMSS by using standard inversion equations (2) or (8) which do not consider $\Theta_1(b,t)$ and $\Phi_1(b,t)$, rather than the CMSS inversion (7) equations, which do. In this case the reconstructed stereo signal will re-mix the spatial concentrations detected by the SISSD to make them center-panned. Doing so allows for the processing of the alternative center signal using any center-panned or center-biased system, including those designed for enhancing dialog in entertainment content such as [20] which targets center-panned speech, or [21] which favors center sources by using an ILD-based mapping. Since these systems were not available for processing private data, we did not pursue these options. (Alternative center signals also allow a simplification of the system proposed in [10] in which the described adaptations for non-center-panned sources become unnecessary.) We plan to explore these ideas in future work.

*Channel Independent*. Perhaps the most obvious alternative processing option is channel independent processing. In this case, depicted in Fig. 2, each of the stereo input channels is processed separately by an SE system. As noted above, for non-center-panned speech mixing where speech exists in both channels, such processing is conceptually similar to processing both standard mid and side signals. The significant difference is for center-panned speech mixing, for which channel independent processing incurs the risk of having different errors in each channel, which may lead to outputs in which listeners perceive distortion. Given that we will use both center-panned and other dialog mixing in the evaluation, we chose to use channel-independent processing as a baseline to allow for meaningful comparisons with CMSS in the greatest number of cases. Processing only the standard mid signal described above, for example, would produce essentially identical results as CMS processing for center-panned speech, preventing meaningful comparison.

**2.4 Alternative Output Format Options**

For the proposed CMS processing system and the baseline channel independent systems, it can be seen that stereo signals are always output, and that the architectures implicitly or explicitly preserve the spatial information of the estimated speech: an input with a center-left speech source should lead to an output with a center-left speech source. We view this as the more challenging case; generating a mono signal, or a trivial stereo signal (e.g. one with both channels identical) does not require the processing system to maintain spatial fidelity. For this reason, we choose to use stereo output for the evaluation.

However, there may be applications for which spatial fidelity is not required, or cannot be included, for example for voice communication systems which only transmit a mono signal to a listener. For CMS-only processing, we may obtain a mono output by having the system directly output the processed custom mid signal rather than use the inversion equations (7). For channel-independent processing, a downmix can be formed. We will consider the quality of mono output signals in future work. For the present, we ask subjects in the evaluation to independently assess speech quality and spatial quality. We will describe more details for the evaluation in the next section.

| Name | SNR | Gender | Dialog Mixing | Genre |
|---|---|---|---|---|
| Outdoors | Mod | F | L-C | News |
| PopMusic | Mod | F | L-C | Ad |
| Bobsled | Low | B | C | Sports |
| Orchestral | Low | F | C | Ad |
| CrowdSing | Low | M | C | Sports |
| Cheering | Low | M | C-R | Sports |
| RaceCars | Low | M | C | Motorsports |
| ShipCrew | Mod | B | C | Scripted |
| Outdoors2 | Mod | M | C | Scripted |
| UrbanSFX | Mod | M | Varies | Scripted |
| SciFiSFX | Mod | F | C | Movie |
| HallDin | Mod | M | Reverb | Movie |

Table 1: Content items for evaluation

## 3 EVALUATION

We now describe subjective listening tests which were used to compare the performance of the two systems described in the previous section: CMS processing and a channel-independent baseline. We describe the test content, processing, methodology and results.

**3.1 Test Content**

For input content, we use stereo items with a variety of dialog mixing styles. As noted in the introduction, we ultimately aim for the proposed system to process content from a variety of potential stereo sources including two-mic mobile devices, two-mic laptops, binaural headphones, and stereo external mics, all of which have device-dependent types of speech mixing. We have done pilot tests of the proposed and baseline systems on stereo inputs captured from known and unknown stereo recording devices, as well as on stereo professionally generated content (PGC), i.e. typical TV and movie content, and found results to be broadly similar in nature for low and moderate SNR content. We attribute this to the generality of the SISSD upon which CMSS are based; it is made to detect spatial concentrations of energy whether they indicate a panned source with some interchannel level difference such as is common in PGC, or a source mixed with interchannel delay or reverb as is expected for environmental capture.

| Attribute | Description |
| --- | --- |
| Preference | Overall preference. Subjects were instructed to identify if they preferred stimulus A or stimulus B. |
| Speech Quality | Identification of speech distortion. Subjects identified which signal had *less* distorted speech. |
| Less Non-Speech | Identification of non-speech sounds. Subjects identified which signal had *less* perceptible non-speech sounds. |
| Spatial Quality | Naturalness of spatial image. Subjects identified which system sounded more spatially natural and in-line with their expectations for a high quality experience. |

Table 2: Attribute descriptions.

For the present evaluation, we use specific PGC items for which the dialog mixing styles can be concisely and accurately described. This allows for evaluation of the proposed and baseline systems for these specific kinds of mixing. Table 1 describes the items by their background type (given as item name), approximate SNR ("Low" indicating less than approximately 5 dB, "Mod" indicating approximately 5-10 dB), speaker gender presentation (where "B" indicates male and female speech in the same item), speech mixing style (C indicating center-panned L-C center-left, C-R center-right) and genre.

### 3.2 Test SE Systems

To create the signals used in the test, we use the processing shown in Fig. 1 and 2. For the SE systems, we use the U-NetFB system described in section 2 above, for both the CMS and baseline conditions. That is, for CMS, one instance of U-NetFB is run on the custom mid signal only. For the channel-independent baseline, two instances of U-NetFB are run, one on each channel. We also did additional listening using DPCRN, with similar structure: for the CMS condition, one instance of DPCRN processed the custom mid signal, and for the channel-independent baseline, two instances of DPCRN were used, one for each channel.

### 3.3 Test Methodology

Nine subjects participated in this experiment and all participants were highly trained in critical evaluation of audio signals. This test was performed in a quiet listening environment with high quality headphones. Subjects were presented with 12 pairwise comparisons in which they evaluated 4 attributes per comparison. In each comparison, subjects were presented with two test stimuli (stimulus A and stimulus B) - each stimulus per trial contained identical source content that was prepared using either CMS or the channel independent baseline (C.I.). The system ordering was randomized across all 12 trials. Subjects were invited to loop sub-sections of the content and to freely switch back and forth between stimulus A and stimulus B in each trial. The 4 attributes under test were overall preference, speech quality, less non-speech, and spatial quality, which are defined in Table 2.

### 3.4 Results

Results are shown in Figures 3, 4 and 5. Figure 3 shows that CMS was selected as the preferred system 73% of the time compared to the C.I. system, across all subjects and all content items. Subjects indicated that CMS processing resulted in better speech quality in 64% of the comparisons under test. Subjects indicated that CMS produced less perceptible non-speech sounds in 72% of the comparisons under evaluation. There was not a notable difference in spatial quality performance across the systems under test. (Additional, less formal listening considering the same content items, conditions and attributes, but with DPCRN [2] instead of U-NetFB for SE processing, found similar audio characteristics.)

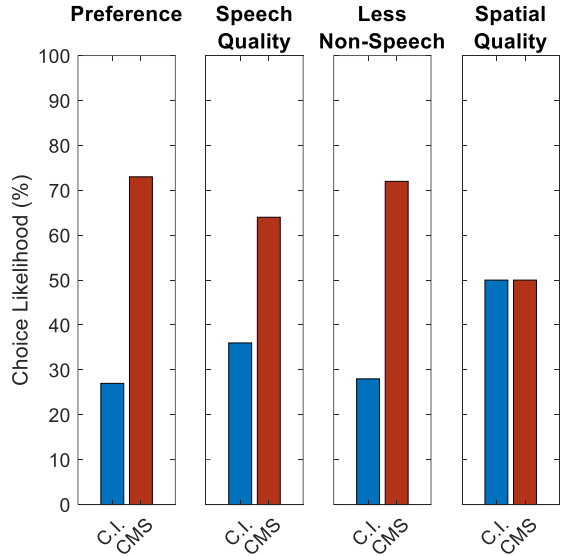

Figure 3: Choice likelihood across all items and subjects for each question.

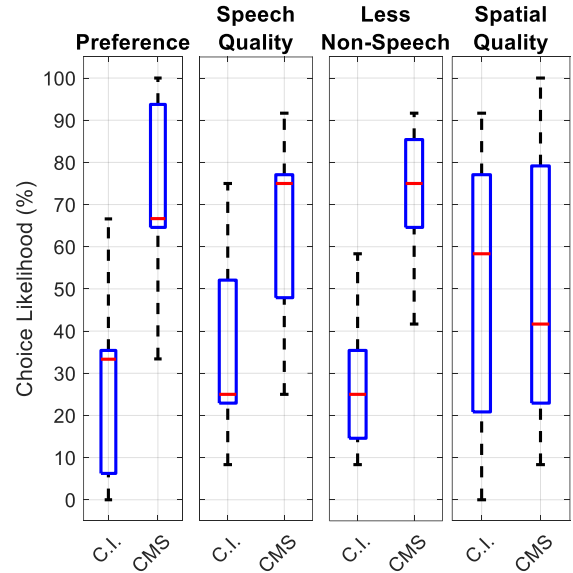

Figure 4: Box plots showing distributions on choice likelihood across all items and subjects for each question.

There was substantial variation between subjects across all attributes under evaluation. Figure 4 depicts the distribution of choice likelihood across all subjects per attribute. This distribution will likely tighten with additional data collection.

Additional analysis found that better speech quality was slightly more correlated to a selection of preference compared to less non-speech and spatial quality (Fig 5). We see a considerably larger correlative range for spatial quality relative to both speech quality and less non-speech indicating that for some subjects spatial quality did not substantially influence a preference outcome. Figures 3 and 4 are consistent with this result. We observe there to be no overall trade-off between preference and processing cost, as the less costly system was also the more preferred system overall.

In future work we will consider how to use existing monaural automated metrics for SE, or novel ones, to evaluate CMS and baseline performance. A pilot investigation found that using existing monaural automated SE metrics to evaluate the stereo CMS and baseline results by averaging the metric values for each channel did not yield results with a meaningful relationship with the subjective data.

## 4 CONCLUSION AND FUTURE WORK

We developed a theory supporting use of custom mid signals as input to any SE system. The theory was based on the idea that the benefits of mid-side signals for center-panned speech sources could be expanded to a much larger class of signals containing speech by using CMSS informed by the SISSD of [10]. An evaluation on various types of speech mixing provided evidence to support our theory, as the CMS processed items were preferred over items processed by the same SE system in a channel independent configuration. This occurred even with the CMS system using only approximately half the computation of the channel-independent system.

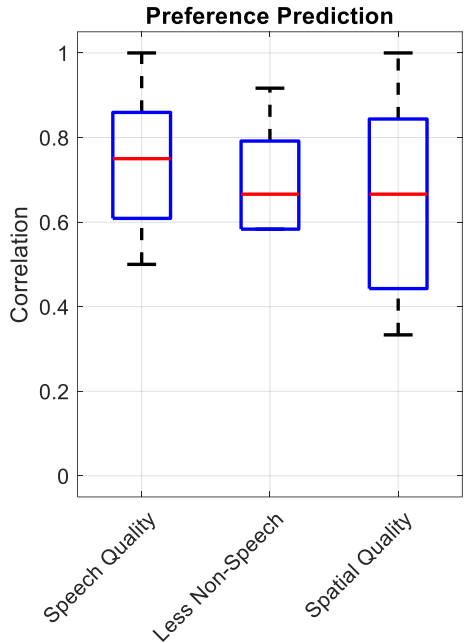

Figure 5: Correlation between questions and preferences across subjects.

In future work, we will perform additional testing of content captured in specific real-world contexts, namely via the two mics on specific, commonly used devices, in specific environments, for applications including UGC capture and voice communications. In particular, we will attempt to find or generate content which contains spatially concentrated non-speech sounds, which we expect could strain the SISSD components of the proposed system [10]. Early pilot testing of UGC content signals from a variety of sources of capture found results that were broadly similar to those on the tested content items for low and moderate SNRs. For all inputs we will investigate whether results can be improved by running SE on both the custom mid and custom side signals.

We noted that existing automated SE metrics did not provide data with a meaningful relationship with the subjective data presented here. We will investigate the underlying causes and consider modifications of these metrics or new ones.

## 5 ACKNOWLEDGEMENTS

The authors thank Heidi-Maria Lehtonen, Scott Norcross, Jonas Samuelsson, Xiaoyu Liu, Dan Darcy, Audrey Howard and Libby Purtill for their assistance with this publication.

## APPENDIX A: STEREO-POLAR DATA

In the CMSS processing system described above, we noted that the SISSD estimates $\Theta_1(b,t)$ and $\Phi_1(b,t)$ are obtained from STFT data. We also noted that the STFT tiles which contain only a particular target source lead to particular estimates. To develop a fuller understanding of these cases, we presently describe a mapping for the data in a typical stereo STFT representation to an alternative form we term *stereo-polar coordinates*. Recall that stereo STFT data is typically represented as the real and imaginary components or magnitude and phase for each channel. The estimation process described in the SISSD first transforms it into an alternative, *stereo-polar coordinates representation* (SPCR) which allows for parameter estimation as described in [10]. When including all four values below, SPCR is convertible to and from a conventional stereo STFT representation, and includes values for the following for each STFT tile, where $L$ and $R$ are the STFT representations of the left and right channels.

$$U = \sqrt{|L|^2 + |R|^2}$$
$$\theta = \arctan\left(\frac{|R|}{|L|}\right)$$
$$\phi = \angle\left(\frac{L}{R}\right)$$
$$\psi = \angle L - \phi \sin^2 \theta$$
$$\phantom{\psi} = \angle R + \phi \cos^2 \theta.$$

The first parameter, $U$, is the *combined channel magnitude* and may be thought of as the data for a combined mono spectrogram. The second parameter, $\theta$, is the *mapped interchannel level difference*, which ranges from 0 (for $L_1$ much greater in magnitude) to $\pi/2$ (for $R$ much greater in magnitude). Unlike alternatives (e.g. Eqn. 22 in [15]) it is strictly bounded. Together, these two quantities describe the *stereo-polar magnitude* of an STFT tile. The third parameter, $\phi$, describes the *interchannel phase difference* from $-\pi$ to $\pi$ radians, and the fourth, $\psi$, the *base phase* also from $-\pi$ to $\pi$ radians; together these describe *stereo-polar phase*. The choice of $\psi$ here with respect to $\phi$ follows a similar convention as for $\Psi_1$ above with respect to $\Phi_1$ in the generalized target signal model. Altogether, we term the four quantities above the SPCR. We can reconstruct $L$ and $R$ from the SPCR by calculating:

$$L = U \cos\theta \exp(i(\psi + \phi \sin^2 \theta))$$
$$R = U \sin\theta \exp(i(\psi - \phi \cos^2 \theta)).$$

We note similarities and differences between the quantities $(\Theta_1, \Phi_1, \Psi_1, |S_1|)$ and $(\theta, \phi, \psi, U)$. The values $|S_1|$ and $\Psi_1$ represent the magnitude and phase of a monaural target source (which can vary across each bin in STFT space), and $\Theta_1$ and $\Phi_1$ are its mixing parameters which characterize its presence in two channels (these parameters vary only if the source moves, is reverberant or mixed with interchannel delay). The values $(\theta, \phi, \psi, U)$, however, are detected for *each* STFT tile, and may or may not coincide with the target source values depending on whether a given STFT tile is dominated by the target source, by interferers, or some combination. (See e.g., [15].) If a given tile is dominated by the target source, it is easy to see by substitution of the $L$ and $R$ values for the generalized source model in section 2 that $(\theta, \phi, \psi, U)$ will equal $(\Theta_1, \Phi_1, \Psi_1, |S_1|)$; this fact forms the basis for the SISSD estimation described in [10], which analyzes distributions on the values $(\theta, \phi, U)$ to calculate $\Theta_1(b, t)$ and $\Phi_1(b, t)$.